# Algorithm to derive shortest edit script using Levenshtein distance algorithm

P. Prakash Maria Liju
ppml38@gmail.com

*Abstract—* **String similarity, longest common subsequence and shortest edit scripts are the triplets of problem that related to each other. There are different algorithms exist to generate edit script by solving longest common subsequence problem. This paper proposes an algorithm that uses string similarity problem to generate shortest edit script. For this we use the famous Levenshtein distance algorithm, which computes a numerical value that represents similarity between the strings from 0 to n, where n is the length of longest input string, and produce the shortest edit script which contains instructions of Insert, Delete and Substitute.**

I. INTRODUCTION

There are many situations where we need to find the shortest edit distance [3] and edit script between different strings or text content. The very common use case is during software development to find the difference between old and new versions of code. Here the edit script contains instructions to insert, delete or substitute characters in source string to convert that into the destination string.

There are different algorithms to derive this edit scripts and one of the famous among them is Myer's algorithm [1]. His algorithms finds the 'difference' between the strings by finding their 'Longest common subsequence' and derive edit script out of it.

One of the widely used application of this algorithm is 'diff' program used in git version control systems, which generates edit script between source codes of different versions.

There is another set of algorithms, that find 'similarity' between the strings or shortest edit distance in other words. One of the famous among these is Levenshtein distance [2] algorithm. This algorithm generates a numeric value that represents the similarity between two strings, where 0 representing same string and higher positive numbers representing non-similar strings in increasing order.

Though this algorithm is widely used in applications like search engine query completion and recommendation engines, Can this be used for shortest edit script generation?

In this paper, we answer this very question, and using Levenshtein distance, we propose an algorithm, that derives shortest edit script between two strings.

II. ALGORITHM PROPOSED

*A. Explanation*

Given two strings A and B, with length m and n respectively, containing characters from a finite space S, our algorithm will derive an edit script E of shortest possible length, which will contain instruction to convert string A to B. These instructions include operations of insertion, deletion and substitution.

Character is taken as the unit of atomicity in this paper for simplicity. This can be scaled to word, sentence, paragraph or file content as per application requirement.

Dynamic programming approach of Levenshtein distance algorithm uses an edit matrix, each cell of which contains minimum edit distance of both the strings till that point. This distance depends on their previous iteration, and the equalness of characters in current iteration. To generate shortest edit script of the strings we need to keep track of all possible edit scripts and select the shortest one. In other words, the script the was generated along the nodes of shortest path in the edit graph from (0,0) to (m, n). As part of Levenshtein distance, since we already maintain edit matrix, along with edit distance, we also maintain shortest edit script till that point, in each cell of the matrix. Then the edit script at cell (m, n) will be the shortest edit script possible.

Suppose, we are given with 2 strings of length m and n, we create an edit matrix of size m+1 * n+1. Variables i and j represent its row and column, starting from 0 to m, n respectively.

At every iteration (i,j), if the characters of strings in current iteration matches, we simply copy the previous iteration (i-1,j-1), else we copy the edit script of the iteration that has minimum distance, (i-1,j) or (i,j-1), and append current edit instruction. If the minimum distance of (i-1,j) and (i,j-1) is equal, we copy (i-1,j-1) instead and append current edit instruction.

Example:

Let's consider "abac" and "aabc" as the two input strings.

|   |   | a | b | c | a |
|---|---|---|---|---|---|
|   | **0** | 1 | 2 | 3 | 4 |
| a | 1 | **0** | 1 | 2 | 3 |
| a | 2 | 1 | **1** | **2** | 2 |
| a | 3 | 2 | 2 | 2 | **2** |

Above is the Levenshtein edit matrix of the strings. Here at cell (1,1) both the strings have same character 'a' at this position, so no edit instruction is required. Whereas (2,2) is a 'substitution' case and (2,3) is a deletion case, as per above explanation.

As we calculate minimum distance, we also identify the edit instruction and we add it along with edit distance in the matrix cell.

*B. Pseudo code*

```
function editScript(char A[1..m], char B[1..n]):
{
// For 0<=i<=m and 0<=j<=n, D[i,j] will hold the Levenshtein distance
// and shortest edit script between
// the first i and j characters of A, B respectively
declare {edit_distance: integer, edit_script: Array} D[0..m, 0..n]
set each element in D to { edit_distance: 0, edit_script: [] }
// Source string can be transformed into empty string
// by dropping all the characters
for j from 1 to m:
{
 D[0, m] := {
 edit_distance: j,
 edit_script: [ "At position j, Delete: A[k]" for every k in [1..j] ]
        }
}
// Target string can be obtained from empty string by adding
// every character
for i from 1 to n:
{
 D[i, 0] := {
   edit_distance: i,
   edit_script: [ "At position i, Insert: A[k]" for every k in [1..i] ]
        }
}
// Calculate shortest edit distance of each cell in the matrix D
// and derive edit script in parallel.
for i from 0 to n-1:
{
 for j from 0 to m-1:
 {
  if A[j] = B[i]:
  {
    // No edit required
    copy D[i][j] into D[i+1][j+1]
  }
  else:
  {
    distance_score := minimum(
       D[i, j+1] + 1,   // deletion
       D[i+1, j] + 1,   // insertion
       D[i, j] + 1      // substitution
       )
   if edit_distance of D[i][j+1] = edit_distance of D[i+1][j]:
     {
     // Substitution
     D[i+1][j+1] := {
      edit_distance:= distance_score,
      edit_script:= Copy edit_script of D[i][j] and append "At position i, Substitute A[j] with B[i]"
           }
     }
     else if edit_distance of D[i][j+1] < edit_distance of D[i+1][j]:
     {
      // Insertion
      D[i+1][j+1] := {
      edit_distance:= distance_score,
      edit_script:= Copy edit_script of D[i][j+1] and append "At position i, Insert B[i]"
           }
    }
     else if edit_distance of D[i][j+1] > edit_distance of D[i+1][j]:
     {
      // Deletion
      D[i+1][j+1] := {
      edit_distance:= distance_score,
      edit_script:= Copy edit_script of D[i+1][j] and append "At position j, Delete A[j]"
           }
    }
   }
  }
}
return edit_script of D[m, n]
}
```

## C. Edit script matrix

Below is how the edit script will be internally represented in the matrix, while input strings are "aabac" and "aabc".

|   |   | a | b | a | c |
|---|---|---|---|---|---|
|   | { edit_distance = 0, edit_script = [] } | { edit_distance = 1, edit_script = ["At position 0 delete a"] } | { edit_distance=2, edit_script=["At position 0 delete a", "At position 1 delete b"]} | { edit_distance = 3, edit_script = ["At position 0 delete a", "At position 1 delete b", "At position 2 delete a"] } | { edit_distance = 4, edit_script = ["At position 0 delete a", "At position 1 delete b", "At position 2 delete a", "At position 3 delete c"] } |
| a | { edit_distance = 1, edit_script = ["At position 0 add a"] } | { edit_distance = 0, edit_script = [] } | { edit_distance = 1, edit_script = ["At position 1 delete b"] } | { edit_distance = 2, edit_script = ["At position 0 delete a", "At position 1 delete b"] } | { edit_distance = 3, edit_script = ["At position 0 delete a", "At position 1 delete b", "At position 3 delete c"] } |
| a | { edit_distance = 2, edit_script = ["At position 0 add a", "At position 1 add a"] } | { edit_distance = 1, edit_script = ["At position 0 add a"] } | { edit_distance = 1, edit_script = ["At position 1 substitute b with a"] } | { edit_distance = 1, edit_script = ["At position 1 delete b"] } | { edit_distance = 2, edit_script = ["At position 1 delete b", "At position 3 delete c"] } |
| b | { edit_distance = 3, edit_script = ["At position 0 add a", "At position 1 add a", "At position 2 add b"] } | { edit_distance = 2, edit_script = ["At position 0 add a", "At position 2 add b"] } | { edit_distance = 1, edit_script = ["At position 0 add a"] } | { edit_distance = 2, edit_script = ["At position 1 substitute b with a", "At position 2 substitute a with b"] } | { edit_distance = 2, edit_script = ["At position 1 delete b", "At position 3 substitute c with b"] } |
| c | { edit_distance = 4, edit_script = ["At position 0 add a", "At position 1 add a", "At position 2 add b", "At position 3 add c"] } | { edit_distance = 3, edit_script = ["At position 0 add a", "At position 2 add b", "At position 3 add c"] } | { edit_distance = 2, edit_script = ["At position 0 add a", "At position 3 add c"] } | { edit_distance = 2, edit_script = ["At position 0 add a", "At position 3 substitute a with c"] } | { edit_distance = 2, edit_script = ["At position 1 substitute b with a", "At position 2 substitute a with b"] } |

Here it can be noticed that edit script is parallelly generated while calculating the minimum edit distance. Hence making the time complexity of this algorithm O(mn)

## D. Time complexity

Our algorithm aligns with the time complexity of Levenshtein distance algorithm. With implementation using dynamic programming approach with a single matrix of m*n size, algorithm will read each and every cell of the matrix calculating its minimum distance and edit script.

Calculation of minimum distance and edit script can be executed in constant time. Hence the time complexity becomes O(1) + O(mn) = **O(mn)**.

## E. Space complexity

Space requirement of our algorithm is equal to the size of edit script matrix. Assuming we use minimal notation for edit instructions (like + for addition – for deletion and ~ for substitution), size of a single instruction becomes constant and the total space complexity is equal to total number edit instructions in the matrix.

In worst case, a cell will contain k number of edit instruction, where k = max (i, j), where i, j are the row and column of the cell in edit script matrix.

Considering below worst case, where the minimum edit sequence is max (m, n).

|   |   | A | b | c | d |
|---|---|---|---|---|---|
|   | **0** | 1 | 2 | 3 | 4 |
| e | 1 | **1** | 2 | 3 | 4 |
| f | 2 | 2 | **2** | 3 | 4 |
| g | 3 | 3 | 3 | **3** | 4 |
| h | 4 | 4 | 4 | 4 | **4** |

Above matrix represents its edit script matrix.

Each cell in above matrix will contain specified number of instructions.
Then total space complexity is,
~= [n*(2n+1)] + [n-1*(2n-2+1)]...+[1*3]+[0]
~= O(n²)
Space complexity of this algorithm is **O(n²).**

## F. Implementation

Python implementation of this algorithm can be found at https://github.com/ppml38/shortest_edit_script

## III. CONCLUSION

In this paper, we presented an algorithm, to derive the shortest edit script using a string similarity algorithm Levenshtein distance, unlike few existing algorithms which use longest common subsequence.